\documentclass[pre,floatfix,showpacs,twocolumn,superscriptaddress]{revtex4-1}
\usepackage{graphicx}
\usepackage{subfigure}
\usepackage{amsmath}
\usepackage{amssymb}
\usepackage{latexsym}
 \usepackage{cancel}
 \usepackage{multirow}
 \usepackage{color}

\begin{document}

\title{
%When fluid mechanics meets elasticity: a geometrical model for liquid ropes.
Liquid ropes: a geometrical model for thin viscous jet instabilities
%When a viscous jets become a rope of fluid \\
%A geometrical model for thin threads fluid mechanical instability. 
%When geometry dominates fluid mechanics instabilites\\
%When fluid mechanics instabilities are driven geometry
%A geometrical model for viscous jets impacting a surface \\
% A geometrical model for the viscous sewing machine \\
}

\author{P.-T. Brun}

\affiliation{CNRS and UPMC Univ.\ Paris 06, UMR 7190, Institut Jean le Rond 
d'Alembert, Paris, France}

\affiliation{Laboratoire FAST, Universit\'e Paris-Sud, CNRS, B\^atiment 502, Campus Universitaire, Orsay 91405, France}

\affiliation{Laboratory of Fluid Mechanics and Instabilities, EPFL, CH1015 Lausanne, Switzerland}

\affiliation{Department of Mathematics, Massachusetts Institute of Technology, Cambridge, Massachusetts 02139, USA}

\author{Basile Audoly}

\affiliation{CNRS and UPMC Univ.\ Paris 06, UMR 7190, Institut Jean le Rond 
d'Alembert, Paris, France}

\author{Neil M.\ Ribe}

\affiliation{Laboratoire FAST, Universit\'e Paris-Sud, CNRS, 
B\^atiment 502, Campus Universitaire, Orsay 91405, France}

 \author{T.\ S.\ Eaves}
 \affiliation{Institute of Theoretical Geophysics, Department of Applied Mathematics and Theoretical Physics,
University of Cambridge, Wilberforce Road, Cambridge CB3 0WA, UK}
 
\author{John R.\ Lister}
 
  \affiliation{Institute of Theoretical Geophysics, Department of Applied Mathematics and Theoretical Physics,
University of Cambridge, Wilberforce Road, Cambridge CB3 0WA, UK}

\date{\today} 

\begin{abstract}

Thin viscous fluid threads falling onto a moving belt behave in a way
reminiscent of a sewing machine, generating a rich variety of periodic
stitch-like patterns including meanders, W-patterns, alternating
loops, and translated coiling.  These patterns form to accommodate the
difference between the belt speed and the terminal velocity at which
the falling thread strikes the belt.  Using direct numerical
simulations, we show that inertia is not required to produce the
aforementioned patterns.  We introduce a quasi-static geometrical
model which captures the patterns, consisting of three coupled ODEs
for the radial deflection, the orientation and the curvature of the
path of the thread's contact point with the belt.  The geometrical
model reproduces well the observed patterns and the order in which
they appear as a function of the fall height. 

\end{abstract}

\maketitle	
%
%\begin{figure}[!h]
%\begin{center}
%\includegraphics[width=.5\textwidth]{phaseD.pdf}
%\caption{}
%%$\mathbf{I}$- DVR simulation of a viscous thread falling onto moving belt : the thread is stretched until a terminal value of the radius $a_c$ associated with the fall speed $U_c$. The lower part of the thread has a "heel" shape and is denoted $\delta$ $\mathbf{II}$ -The heel $\delta$ lays down on the belt forming sewing machine patterns. $\mathbf{III}$- Numerical phase diagram obtained with DVR. Translated coiling (red online), alternating loops (green online), meanders (blue online), and W-pattern (orange online) are shown in the parameters plane $( V,  H)$. $\mathbf{IV}$-The later diagram is advantageously rescaled by $U_c$. Note the horizontal boundaries and the reported hysteresis between them.}
%\label{phaseD}
%\end{center}
%\end{figure}

\begin{figure}[b!]
    \begin{center}
	\includegraphics[width=.5\textwidth]{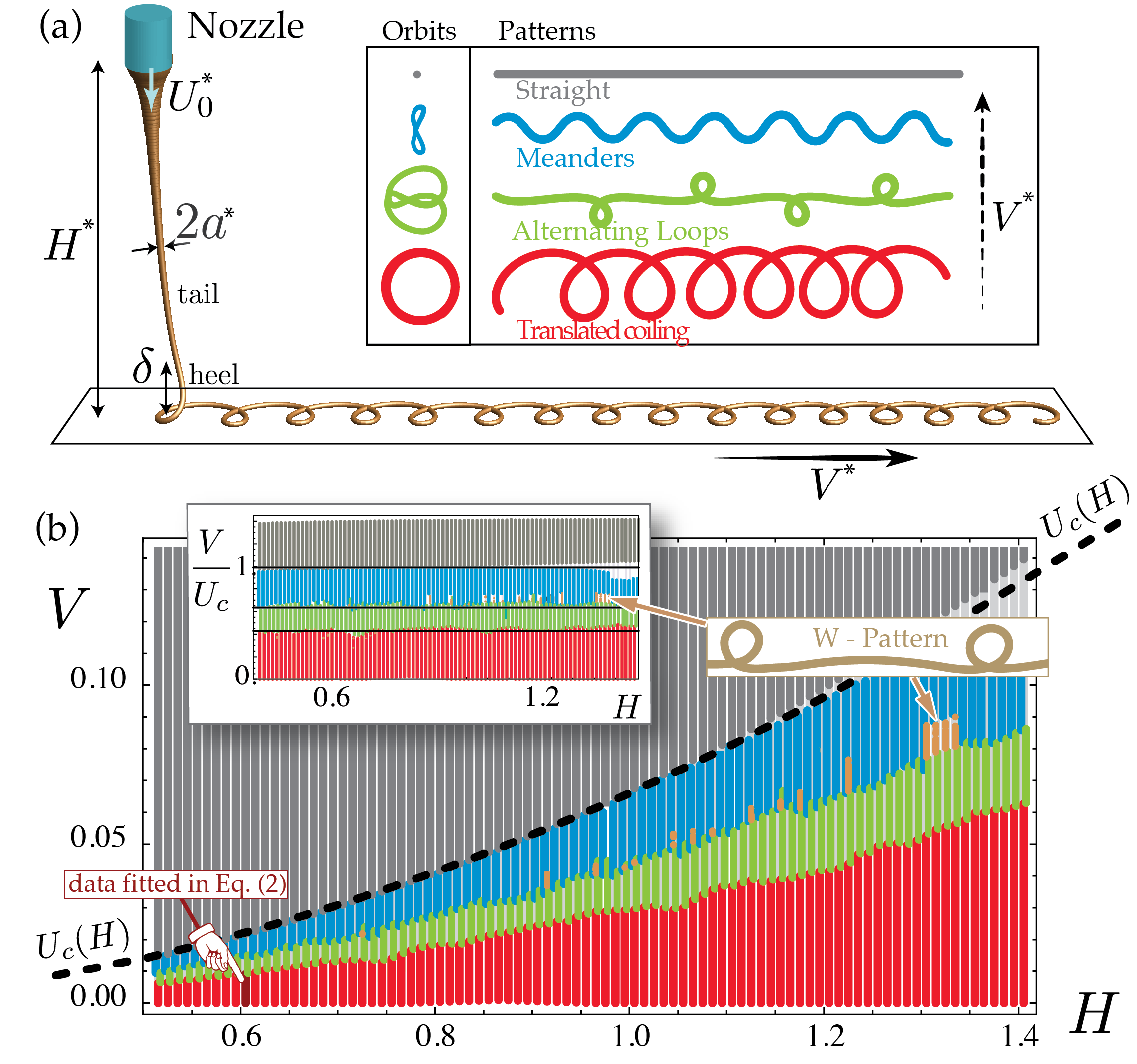}
	\caption{(a) Direct numerical simulation, with no inertia, of a thin thread of
	viscous fluid falling from a height $H^*$ onto a belt of
	velocity $V^*$.  Shown are four periodic orbits of the contact
	point of the thread on the belt, and the corresponding spatial
	patterns.  (b) Phase diagram showing the distribution of
	patterns in the \emph{dimensionless} parameter plane $( H, V)$.  The
	speed $ U_c(H)$ at which the fluid coils in the absence of
	advection (${V} = 0)$ is shown by the dashed line.  Inset:
	same diagram, with belt velocity rescaled by coiling velocity
	$ U_c(H)$. }
	\label{phaseD}
    \end{center}
\end{figure}

A thin thread of viscous fluid falling onto a moving belt is a
remarkable pattern-forming system with surprisingly complex behavior.
The patterns laid down onto the belt include meanders, alternating
loops, W-pattern, coiling (Fig.~\ref{phaseD}),
as well as various
resonant patterns such as double coils and double
meanders~\cite{ChiuWebster:2006jd,Morris}.  The resemblance of these
patterns to the stitch patterns of a sewing machine led
\cite{ChiuWebster:2006jd} to call the system the ``fluid mechanical
sewing machine" (FMSM).  The FMSM is of interest as a simplified model
for industrial processes such as the production of non-woven
textiles~\cite{Marheineke:2009kp} and the laying down of ``squiggles"
of icing on cakes.  It is also an accurate model for one of the
characteristic gestures of Jackson Pollock's action painting, in which
paint from a moving brush dribbles onto a stationary horizontal
canvas~\cite{Herczynski:2011ws}.  But the FMSM also has fundamental
interest as an example of great complexity (roughly a dozen distinct
patterns) arising in an extremely simple system (a single thread of
Newtonian fluid).  This fundamental interest has inspired several
experimental, theoretical and numerical studies of the FMSM in recent
years~\cite{ChiuWebster:2006jd,Morris,Audoly:2012wk,Brun:2012ic}.

The FMSM patterns are best thought of as resulting from the lateral
advection of the periodic orbits (in the frame of the nozzle) of the
thread's contact point with the belt.
%These orbits may well be characterized by their frequency spectrum which combines exact ratios of the natural frequency of the thread denoted $\Omega_c$ as it is the frequency at which the thread coils when falling onto a fixed  substrate. 
The canonical example of such a periodic orbit is the circular orbit
produced by the steady coiling of a highly viscous fluid thread (e.g.,
honey) falling onto a surface (e.g., toast).  It has been observed
that the frequencies of the FMSM pattern are all simple multiples of
the steady coiling frequency $\Omega_c$~\cite{Brun:2012ic}.  In this
Letter we reveal the unforeseen physical mechanism underlying this
result, and show that it differs fundamentally from the typical
harmonic resonance in parametric oscillators such as Mathieu's 
which, in terms of frequency content, has strong similarities
with the alternating loop pattern~\cite{Brun:2012ic}.  The novel aspect of the system we study
resides in the fact that no inertia is needed to produce the patterns.
Accordingly, we account for the observations using a simple
three-variable dynamical system for the radial deflection, the
orientation and the path curvature of the contact point.

Before deriving the model we perform direct simulations of the FMSM  with the Discrete Viscous Rods algorithm (DVR)~\cite{Audoly:2012wk,Brun:2012ic} to propose a rationalization of the FMSM phase diagram when inertia is negligible. 
%The DVR algorithm is based on a discrete center-line description of the viscous thread and makes use of Rayleigh potentials for describing stretching, bending and twist~\cite{Audoly:2012wk}.  It allows for unsteady simulations of viscous threads such as required in the case of the FMSM~\cite{Brun:2012ic}.
Consider a thread with kinematic viscosity $\nu$ falling at a
volumetric rate $Q^*$ from a nozzle of dimensional height $H^*$ onto a
conveyor belt moving horizontally at speed $V^*$.  The thread is
stretched by gravity (denoted $g$) during its fall so that the speed
of the fluid increases with distance from the nozzle.  Balancing the
gravitational stretching with the viscous dissipation in the thread
yields a typical length scale $(\nu^2/g)^{1/3}$ and time scale
$(\nu/g^2)^{1/3}$ that we use to nondimensionalize our equations.  In
particular, $H=H^*(g/\nu^2)^{1/3}$ and $V=V^*/(\nu g)^{1/3}$ are the
dimensionless height of fall and the dimensionless belt velocity
respectively.  Varying these two parameters independently allows us to
generate a phase diagram for the FMSM~\cite{Brun:2012ic}.  In such a
diagram, inertial effects are generally negligible when working with
physical parameters such that $ H\ll1$.  Herein, we make the choice of
working with the typical parameter values used in the
literature~\cite{Morris} such that $0.5\leqslant H\leqslant1.4$, but
we artificially omit inertia in our numerical simulations.  The
purpose is to identify the patterns which survive this quasi-static
limit and show that nonlinearities in this system are independent of
inertia.  The patterns (Fig.~\ref{phaseD}) which we found are detailed
next.

%The purpose is to compare the newly produced inertialess phase diagram to the one which would have been produced in normal conditions~\cite{Brun:2012ic}, that is to probe the role of inertia in the pattern formation.  
%In facts, non-trivial patterns are reported  in the quasi-static limit as detailed in the subsequent paragraph. 
%Those 'inertialess' patterns are now described operating at a fixed height of fall $ H$  while increasing the velocity of the belt $ V$ in the phase diagram shown in Fig.~\ref{phaseD}.

When the belt has velocity $ V=0$ the thread coils steadily with a
radius $ R_c$, frequency $ \Omega_c$ and speed $U_c=R_c\Omega_c$
(steady coiling) ~\cite{Ribe:2012uo}.  When gradually increasing the
belt velocity while keeping other parameters constant, the coiling
pattern is first simply translated on the belt (translated coiling) up
to a certain critical value of $V$ where loops form alternatively on
one side of the belt and then the other (alternating loops).  For
higher belt speeds the thread exhibits some
meanders~\cite{Ribe:2006gz,Blount:2011ke} which collapse to a straight
line for a critical value of the belt velocity $ V_c$.  For velocities
higher than $ V_c$ the thread has a catenary shape and its contact
point with the belt is stationary in the laboratory frame.  In the
rest of the Letter we concentrate on belt speeds in the range
$0\leqslant V\leqslant V_c$.  Three points are of particular interest.
First, no double patterns~\cite{Brun:2012ic} such as the double
coiling or double meanders were found in these quasi-static
conditions.  This was anticipated since such resonant patterns are
typically observed for large values of $ H$ where inertia is dominant
in normal conditions~\cite{Brun:2012ic}.  Second, we found hysteresis
in the critical belt velocity values corresponding to the transition
between patterns.  The data shown in Figure~\ref{phaseD}b correspond
to the situation where the belt was slowly accelerated.  The case of a
decelerating belt is discussed at the end of the Letter.  Third, we
report the presence of another pattern --- the W-pattern --- which we
found in limited portions of the diagram (see overlay in
Fig.~\ref{phaseD}b).  It appears in competition with the meanders
after the alternating loops become unstable when the belt speed is
increased (and only then).

\begin{figure}[!h]
\begin{center}
\includegraphics[width=.48\textwidth]{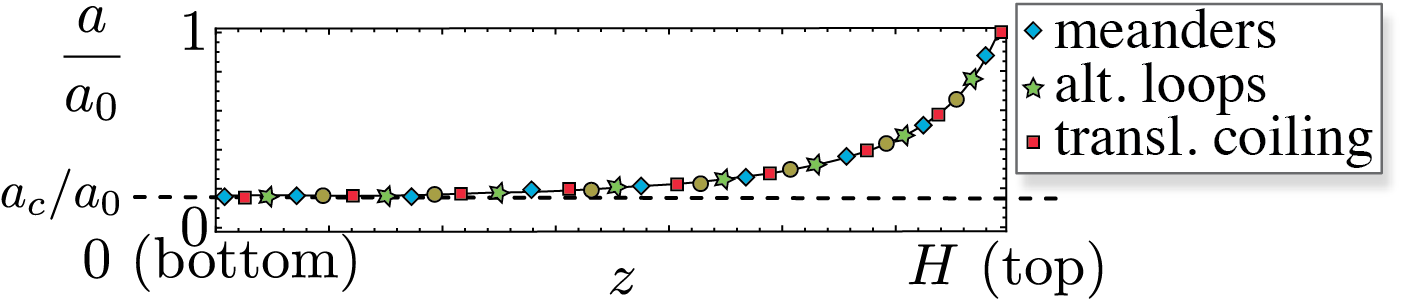}
\caption{The thread's radius distribution $a(z)$, here normalized by
the nozzle's radius $a_{0}$, is the same for any pattern in the range
$ V< V_c$. Stretching occurs in the upper part of the thread so that the radius of the thread is 
approximately uniform in the vicinity of the belt. }
\label{phaseD2}
\end{center}
\end{figure}

For any height ${H}$, we can compute the steady coiling velocity
$ U_c\equiv  R_c\Omega_c$ using the method
of~\cite{Ribe-Coiling-of-viscous-jets-2004}.
% (note that in dimensional form $ U_c$ scales as $gH^2/\nu$), 
This yields the dashed curve in Fig.~\ref{phaseD}b.  The curve matches
the lower boundary of the grey region (straight pattern), which
reveals that the onset of steady coiling matches accurately the critical
velocity $ V_c = U_c$.  The central role played by the reduced
velocity $ V / U_c$ in the formation of the patterns becomes
even more evident when one plots the phase diagram in terms of
${V}/{U}_{c}$, see inset in Fig~\ref{phaseD}b: then, all
boundaries between patterns become horizontal straight lines.  This
important finding shows that the only influence of the height of fall
on the patterns is to set the value of the reduced velocity
${V}/{U}_{c}({H})$: the patterns can be rationalized
strictly in terms of the parameter ${V}/{U}_{c}({H})$.

The reason why ${V}/{U}_{c}$ is the only relevant parameter may be
understood by analyzing the thread's radius profile $a(z)$ for
different $ V$ while keeping $ H$ constant, \textit{i.e.}\ moving
vertically in the phase diagram and browsing through the different
patterns.  We do so in Fig~\ref{phaseD2} and find that all the curves
$a(z)$ collapse onto a single master curve.  In the upper part of the
master curve, called the tail, the thread is accelerated and stretched
by gravity until it reaches a terminal radius $a_c$.  Both this radius
and the speed $Q/(\pi {a}_c^2)$ at which the thread arrives on the
belt are found to be approximately independent of $V$ in the range
$0\leqslant V \leqslant V_c$.  As a consequence the thread speed may
be called the free-fall speed~\cite{ChiuWebster:2006jd} and is equal to
the coiling speed $U_c$ (observed when $V=0$) which solely depends on
$H$.  In general $U_c$ and $V$ do not match and there is a small
region near the lower end of the thread, called the heel in
Fig.\ref{phaseD}, where the thread bends and twists while keeping a
constant radius.  The patterns are produced as the heel is set in
motion to satisfy the no-slip boundary condition at the contact point
between thread and belt:
\begin{equation}
    U_{c}\,\mathbf{t} + V\,\mathbf{e}_{x} = \dot{\mathbf{r}}
    \label{clamp}
\end{equation}
\begin{figure}
    \begin{center}
	\includegraphics[width=\columnwidth]{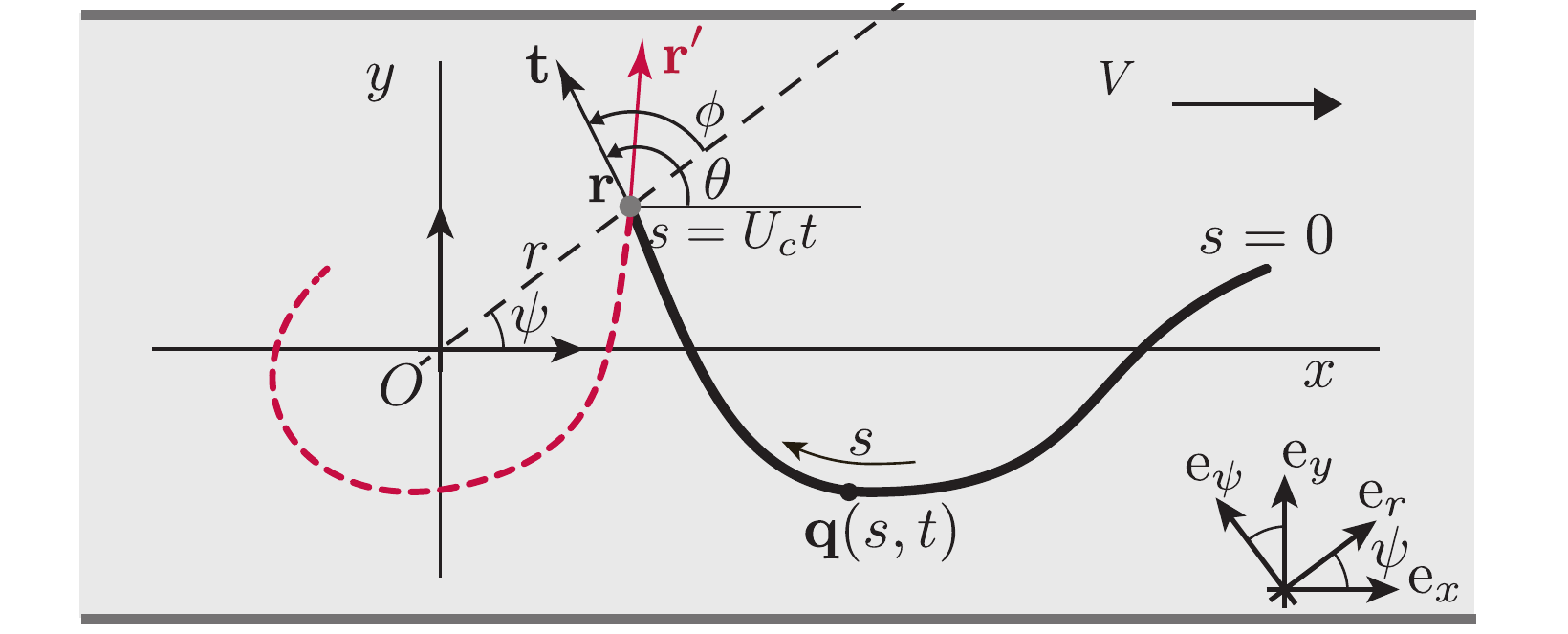}
	\caption{Sketch of the geometrical model in the plane of the belt:
	deposited trace $\mathbf{q}$ (thick black curve) parameterized
	by its arc-length $s$, orbit of the contact point
	(dashed red curve).  The curvature of the thread is assumed to
	be a function of the polar coordinates $(r,\phi)$ of the point
	of contact $\mathbf{r}$.  The projection $O$ of the nozzle
	onto the belt's plane is used as the origin.}
	\label{fig:drawkappa}
    \end{center}
\end{figure}%
Here we use the notation introduced in Fig.~\ref{fig:drawkappa}:
$\mathbf t$ is the unit tangent to the thread at the point of contact
$\mathbf{r}$ with the belt, $\dot{\mathbf{r}}$ is the velocity in the
laboratory frame of this non-material point, and $\mathbf{e}_{x}$ is a
unit vector in the direction of belt motion.  The limiting case of
steady coiling corresponds to $ V = 0$ and $\dot{\mathbf{r}} =
 U_c\,\mathbf t$, and the case of a straight
(catenary) pattern corresponds to $\dot{\mathbf{r}} = \mathbf{0}$,
$\mathbf{t} = -\mathbf{e}_{x}$ and ${V} = {U}_{\mathrm{c}}$.
In the general case $ V/ U_c<1$, the speed at which the thread arrives at
 the belt exceeds the belt's ability to carry it away in a
straight line ($\dot{\mathbf{r}} \neq \mathbf{0}$ in equation above).
This excess length of thread is accumulated on the belt in the form of
patterns produced as the heel lays down on the belt.  This agrees with
our initial observation that the critical velocity at which the
straight pattern appears is $ V_c = U_c$, see
Fig~\ref{phaseD}b.

\begin{figure}
    \begin{center}
	\includegraphics[width=.9\columnwidth]{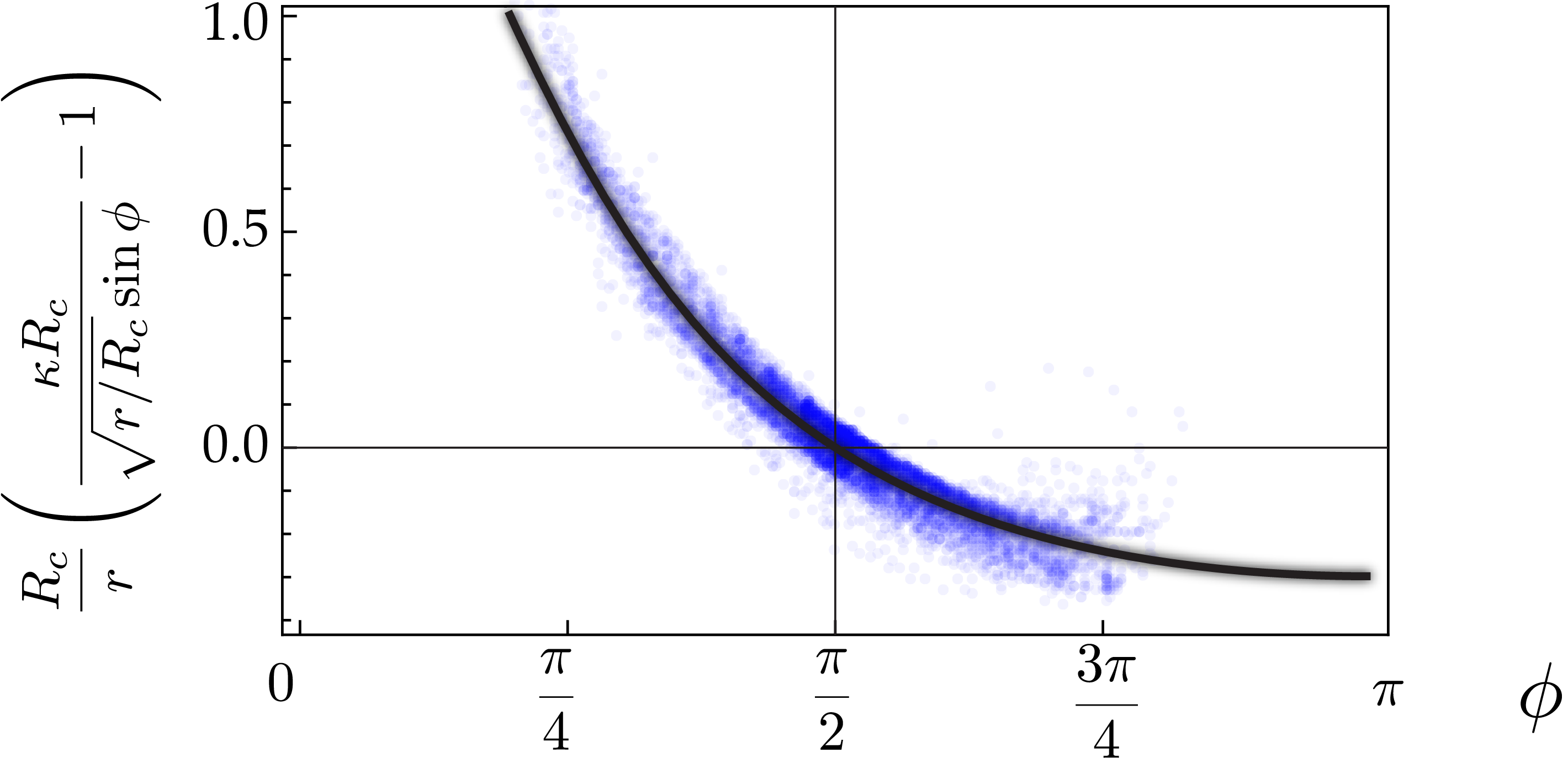}
	\caption{Collapse of the DVR simulation data for the rescaled
	curvature as a function of $\phi$, for the translated coiling
	pattern ($H = 0.6 $ and $0< V/ U_c < 0.4$, see darker red bar
	in the lower left corner of figure~\ref{phaseD}b).
% 	$r$ varies in the range $0.18 \leqslant r \leqslant 1.92$.
	See Supplementary Information for details.}
	\label{mastercurve}
    \end{center}
\end{figure}

We now turn to the task of characterizing and then modeling the heel
boundary layer where the deposition takes place.  Since bending
stresses are dominant in the heel, we anticipate that the curvature
$\kappa$ of the thread at the point of contact plays a key role in the
pattern formation.
Working in the quasi-static (inertialess) limit,
we assume that the shape of the hanging thread (and in particular its
curvature near the point of contact) is only a function of the current
boundary conditions applied to the thread.
% 
% Because the patterns have shown to be closely related to the stated
% coiling problem we now use $R_c$, the coiling radius, to
% nondimensionalize our equations.  Doing so, we automatically capture
% the influence of the height of fall $H$ and seek to reveal the generic
% mechanism leading to the horizontal boundaries in the state diagram,
% see inset in Fig~\ref{phaseD}b.
% 
The boundary conditions at the nozzle are time-invariant as the fall
height and flow rates are fixed.  Therefore, we view the curvature
$\kappa$ at the bottom of the hanging thread as a function of the
position $\mathbf{r}$ of the point of contact and the orientation of
the tangent $\mathbf{t}$.  The equations for the hanging thread are
cylindrically invariant, and therefore we have $\kappa =
\kappa(r,\phi)$, where $\phi$ is the direction of the tangent relative
to the line joining the projection of the nozzle $O$ to the point of
contact $\mathbf r$ (Fig.~\ref{fig:drawkappa}).  The function
$\kappa(r,\phi)$ is found by fitting DVR simulations of translated
coiling for the case $ H = 0.6 $ and $0< V/ U_c < 0.4$ (darker red bar in the
lower left corner of figure~\ref{phaseD}b).  As explained in the
Supplemental Information, time series of $(r,\phi,\kappa)$ for the
translated coiling pattern are well approximated by the heuristic fit
% \begin{align} 
\begin{equation}
    \kappa(r,\phi) =\frac{1}{R_c}\,\sqrt{\frac{r}{R_c}}\,
    \left(1 + A(\phi)\,\frac{r}{R_c}\right)\,\sin\,\phi
    \label{eq-toyeq-kappa}
\end{equation}
%     \\
%     \intertext{where}
%     A(\phi) = \left(b^{-1} - \cos\phi\right)^{-1}-b,\quad
%     b = 0.715
%     \textrm{.}
%     \label{eq-toyeq-A}
% \end{align}
where $A(\phi) =b^2 \cos\phi/(1-b\cos\phi)$ and $b =
0.715$ and $R_c$ is the radius of steady coiling~\cite{Ribe:2012uo}.  Fig~\ref{mastercurve} shows the collapse of the numerical
data obtained from Eq.~(\ref{eq-toyeq-kappa}).
% This expression of the curvature will be used in the geometrical model
% hereafter derived.  
% Note that the fitting function $\kappa$ in Eq.~(\ref{eq-toyeq-kappa}) 
% has been obtained from a tiny region of the phase diagram 

Building on our previous observations, we now derive a quasi-static
geometric model for the formation of the trace.  The heel is modeled
as a filament of uniform radius falling towards the belt at a velocity
$U_c$, which is bent and laid down quasi-statically onto the belt.
Let $s$ be the arc-length along the trace, with 
$s=0$ corresponding to the point which contacted the moving belt at
time $t=0$ and $s=U_c t$ corresponding to the current point of
contact $\mathbf r$. We label
material points in the trace by their (Lagrangian)
coordinate $s$. We also use $s$ as a time-like variable and write $\mathbf{r}(s)$ for the contact position at time $t=s/U_c$.
Let $\mathbf{q}(s,t)$ be the position on the belt of the point $s$ at
time $t$, with $0\leq s \leq U_c t$.  This point was deposited at time
$ s/U_{\mathrm{c}}$ at position $\mathbf r(s)$, and has
subsequently been advected at velocity $V\mathbf{e}_x$ by the belt. 
Thus
\begin{equation}
    \mathbf{q}(s,t) = \mathbf{r}(s) + 
    V\,(t-s/U_c)    \,\mathbf{e}_x
    \textrm{.}
    \label{eq:advect}
\end{equation}
In our model of the thread, the dynamical quantities of interest are the contact position $\mathbf r$, and
the tangent vector $\mathbf t$ and curvature $\kappa$ at the point of contact.
  At any point $s$, the tangent 
to the trace is  $\partial \mathbf{q}/\partial s$.  In
particular, at the point of contact $\mathbf{t}(s) = 
\left.\partial \mathbf{q}/\partial s\right|_{s=U_{c}\,t} =
\mathbf{r}'(s) - V/U_{\mathrm{c}}\, \mathbf{e}_{x}$, and we recover
Eq.~(\ref{clamp}) with $\mathbf{r}' = \dot{\mathbf{r}}/U_{c}$. 
Now let $(r(s),\psi(s))$ denote the polar coordinates of the contact point
$\mathbf{r}(s)$
as shown in Fig.~\ref{fig:drawkappa}, and let $\theta(s)$ denote the angle from the
$x$-axis to $\mathbf{t}(s)$. 
We resolve $\mathbf r'$, $\mathbf{t}$ and $\mathbf{e}_x$ into the polar basis
$(\mathbf{e}_{r},\mathbf{e}_{\psi})$, and use $\phi = \theta-\psi$ to eliminate the
dependence on $\phi$:
%We write the equation for $\mathbf r'$ in the polar basis
%$(\mathbf{e}_{r},\mathbf{e}_{\psi})$, and use $\mathbf{t}=(\cos\phi,\sin\phi)$ with $\phi = \theta-\psi$ to eliminate the
%dependence on $\phi$:
%
%Noting that $\phi = \theta-\psi$, we
%write the equation for $\mathbf r'$ in the polar basis
%$(\mathbf{e}_{r},\mathbf{e}_{\psi})$, thereby eliminating all
%dependence on $\phi$:
\begin{subequations}
    \label{eq:toyeq}
\label{eq:toyeq-2}
\begin{align}
    r' & = \cos(\theta-\psi) + 
    \frac{V}{U_{\mathrm{c}}}\,
    \cos\psi
    \label{eq:toyeq-RPrime}\\
    r\,\psi' & = \sin(\theta-\psi)-
   \frac{V}{U_{\mathrm{c}}}\,
    \sin\psi
    \textrm{.}
    \label{eq-toyeq-PsiPrime}
\end{align}
Finally, $\theta'$ is the curvature of the trace at the contact point,
which has been found in Eq.~(\ref{eq-toyeq-kappa}) in terms of a
fitting function $\kappa$:
\begin{equation}
    \theta' = \kappa(r,\theta-\psi). 
    \label{eq-toyeq-Curvature}
\end{equation}
\end{subequations}

\begin{figure}
    \begin{center}
	\includegraphics[width=.5\textwidth]{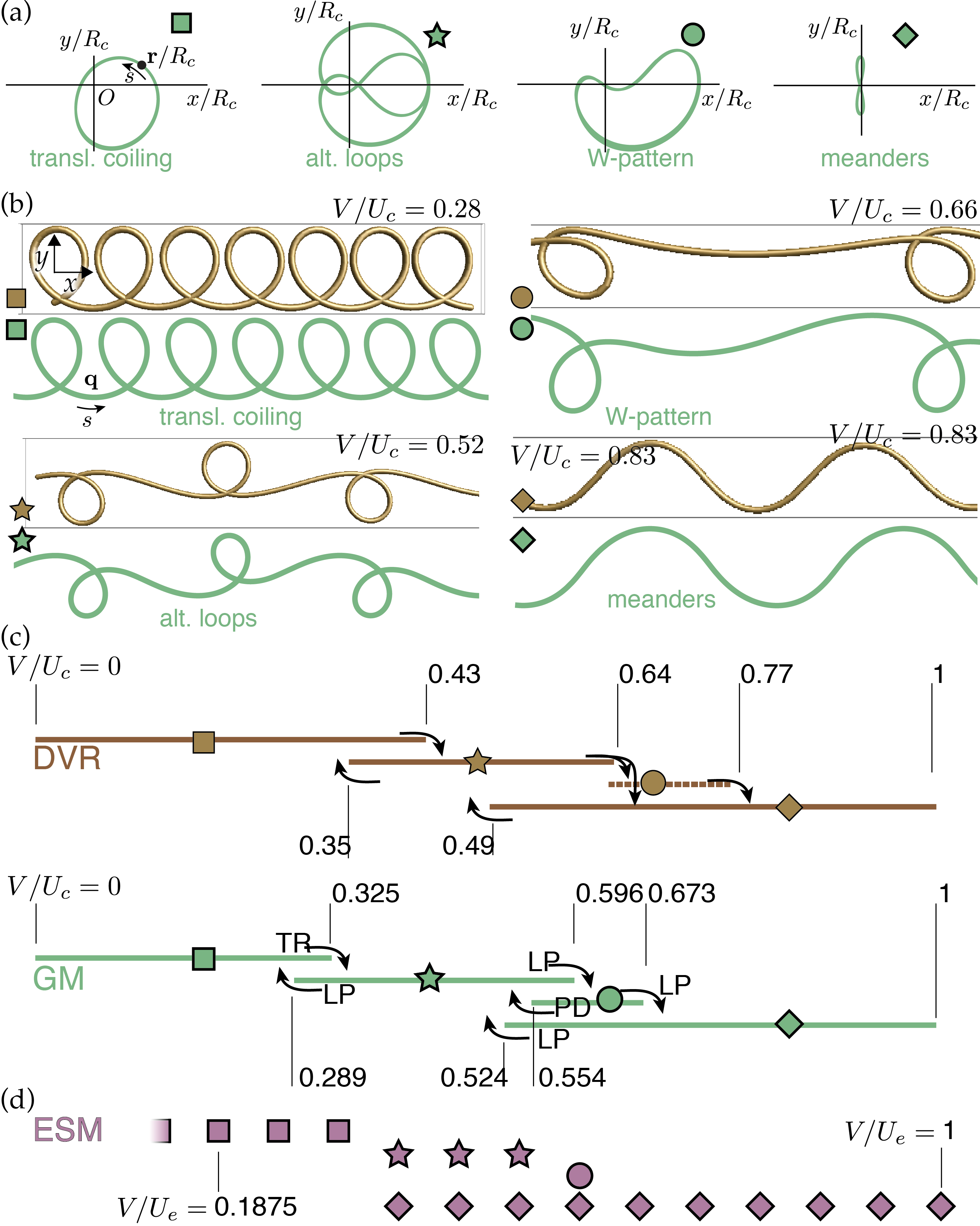} 
	\caption{ (a) The four periodic orbits $\mathbf{r}(s)$
	obtained with the GM and (b) the corresponding patterns
	$\mathbf{q}(s,t)$ (green), compared to the pattern obtained
	with DVR simulations (brown) for identical ratios $V/U_c$.
	(c) Patterns encountered with DVR while quasi-statically
	increasing the ratio $V/U_c$ (resp.\  decreasing, as indicated
	by the arrows) along with the stability domains and
	bifurcation analysis computed with the GM (green): period
	doubling (PD), fold point (FP), torus bifurcation (TR).  (d)
	Experimental patterns for the \emph{elastic} sewing machine,
	from~\cite{Habibi:2011jr}.}
	\label{results}
\end{center}
\end{figure}

Equations~(\ref{eq:toyeq-RPrime}--\ref{eq-toyeq-Curvature}) are a set
of coupled ordinary non-linear differential equations for the
functions $r=r(s)$, $\psi=\psi(s)$ and $\theta = \theta(s)$, depending
on a \emph{single} dimensionless parameter ${V}/{U}_{c}$ --- the
parameter $R_{c}$ in equation~(\ref{eq-toyeq-kappa}) sets a
lengthscale for $r$ and $s$, and can be removed by rescaling.  We
refer to this system of differential equations as the
\emph{geometrical model} (GM).  The kinematic
equations~(\ref{eq:toyeq-RPrime}--\ref{eq-toyeq-PsiPrime}) capture the
coupling with the moving belt, while
equation~(\ref{eq-toyeq-Curvature}) captures the shape of the hanging
thread as set by the balance of viscous forces and gravity.  We
integrated the GM numerically, varying the velocity parameter in the
range $0\leqslant {V}/{U}_c \leqslant1$ (Fig~\ref{results}).  The
solutions $\mathbf{r}(s)$ were found to settle into periodic orbits,
see Fig.~\ref{results}a.
The patterns corresponding to the different orbits can be identified
by reconstructing the complete trace $\mathbf{q}$ from
Eq.~(\ref{eq:advect}), and then compared to those obtained by DVR
simulations, see Fig.~\ref{results}b.  With the aim to calculate the
bifurcation thresholds accurately and to identify the nature of
the bifurcations, we also investigated the stability domains of the
periodic solutions of the GM using the continuation software AUTO
07p~\cite{Doedel}, see Fig.~\ref{results}c.

All the patterns originally observed with DVR in the quasi-static
(non-inertial) limit are captured by the GM. They appear in the
correct order when $ V/ U_c$ is varied, and there is a good agreement
on the values of the pattern boundaries, see Fig.~\ref{results}c.
Their shapes are accurately captured as well, see Fig.~\ref{results}b.
Alternating loops and meanders are symmetric about $y = 0$ in their
full domain of existence, both in DVR simulations and in the GM. The
alternating loops, and the amplitude of meanders both decrease as the
belt velocity increases and the latter tends to zero when $ V= U_c$,
as expected.  Coils are symmetric at zero belt velocity, but then turn
asymmetric at larger velocities.  W-patterns are, on the other hand,
always asymmetric.

%Note that the points of the teardrop of the
%W-patterns become elongated at the lower end of their range of
%stability and become much flatter (though they do not disappear
%entirely) at the upper end. 

% This is quite remarkable as the heuristic fit for the
%curvature has been realized using only data from translated coiling,
%hence on a limited range of values for $V/U_c$.  Nevertheless, the
%model remains coherent for all values of $V/U_c$ and even provides a
%reasonable set of belt velocities for the transitions from one pattern
%to another as seen in Fig.~\ref{results}c.

Interestingly, the GM sheds light on two subtle features of the FMSM.
First, it accounts for the hysteresis observed in DVR when 
transitions between patterns occur at different values depending on
whether the belt velocity is increasing or decreasing: the domains of
stability of the various patterns predicted by the GM do indeed overlap,
see Fig.~\ref{results}c.  Second, it explains why the W-pattern can be
observed in DVR with an increasing belt velocity, but not with a
decreasing one.  Indeed, the layout of the stability diagram of the GM
in Fig.~\ref{results}c predicts that meanders will destabilize
directly into alternated loops for a decreasing belt velocity,
skipping the W-pattern.

% the W-pattern domain of stability is
% found to be strictly included in the stability domain of meanders
% (Fig.~\ref{results}c).  Therefore, the W-pattern is always in
% competition with meanders, since both are stable in the range $0.554<
% V/ U_c<0.673$.  The selection mechanism favoring the meanders over the
% W-pattern (as found most of the time with DVR) is yet to be
% determined.  Nevertheless, our results confirm (a) that W-pattern is
% not necessarily seen while the belt velocity is slowly increasing and
% (b) it is never observed when the belt velocity is decreasing (as seen
% with DVR).

%Another interesting feature of the FMSM is the frequency content of the orbits from 
%Fig~\ref{results}a. In most cases, frequencies are locked on
% simple ratios of the coiling frequency $\Omega_c$~\cite{Brun:2012ic}. In particular, the alternating
% loop pattern corresponds to the combination of the first five multiples of $\Omega_c/3$.
% This feature is reminiscent of parametric oscillators such as Mathieu's whose  resonant modes write $n\,\Omega_0/2$ where $\Omega_0$ is the natural frequency of the oscillator. Herein, the novelty resides in the fact that such a combination is recovered with the GM, that is without inertia but instead taking into account the position and the tangent (that is the orientation) to the thread at the contact point with the belt.  

We now examine the relevance of the GM to the \emph{elastic} sewing
machine, made up of an elastic thread fed downwards onto a belt moving
at a speed $V$.  This system has been analyzed by a combination of
methods including experiments, direct numerical simulations, and
scaling arguments~\cite{Habibi:2011jr,%
Bergou-Wardetzky-EtAl-Discrete-Elastic-Rods-2008,%
Jawed-Da-EtAl-Coiling-of-elastic-rods-2014}.  The main difference
between the elastic and viscous systems resides in the thread rheology.  In particular,
the elastic thread is not stretched by gravity, and the role assigned
formerly to the coiling velocity $U_{c}$ is now played by the feeding
velocity $U_{f}$.  This suggests that $V/U_f$ should be used as the
control parameter in the elastic case.  In terms of this parameter, we
summarize in Fig.~\ref{results}d the experimental results
of~\cite{Habibi:2011jr} for a small height of fall (that is when
inertia is negligible), and compare with the behavior of the FMSM.
Even though their constitutive laws are fundamentally different, the
two systems produce the same set of patterns, and these appear in the
same order and at comparable values of their respective control
parameter.
% 
% 
% 
% 
% In fact, both viscous and elastic threads have a rotational symmetry,
% preserved when buckling in the steady coiling regime, but broken by
% the belt which travels in the direction $\mathbf e_x$ to leave a
% preserved reflection symmetry in a vertical plane containing the axis
% $y=0$ (weakly broken $O(2)$ symmetry). 
As a possible explanation to these similar behaviors, we note that
both systems feature a weakly broken $O(2)$ symmetry~\cite{Morris},
\emph{i.e.}\ the cylindrical symmetry corresponding to steady coiling
is broken by the belt travelling in the direction $x$, to leave a
mirror symmetry with respect to the vertical plane $(Oxz)$.

Returning to the fluid-mechanical sewing machine, we note that the
geometrical model is based on a fit of the curvature in a
very small portion of the phase diagram (see darker bar in the lower
left-hand corner of figure~\ref{phaseD}), yet it successfully predicts
the entire phase diagram.  This, together with the fact that the
patterns are almost identical in the elastic case, points to the key
role played by geometry.  The geometrical model is formulated as an
evolution problem for the position of the contact point, with an
additional dependence on the tangent orientation; compared to a
standard oscillator, this dependence induces a memory effect which explains the
complexity of the pattern, even in the absence of inertia.

% To conclude we recall the geometrical novelty of the dynamical system,
% that is a two-dimensional oscillator for the position of the contact
% point, with an additional variable associated with the orientation
% giving to the system a complex behavior even though there is no
% inertia.  In particular, this low-order system successfully describes
% the complex dynamics of a fully three-dimensional time-dependent flow
% despite its simplicity.

% \bibliographystyle{unsrt}	
%\bibliography{GeometricalModelMS}	

\end{document}